\documentclass[letter]{jpsj3}
\usepackage{scalefnt}
\usepackage{txfonts}
\usepackage[dvipdfmx]{graphicx}
\usepackage{color}

\title{Variation of Pressure-Induced Valence Transition with Approximation Degree in Yb-Based Quasicrystalline Approximants}

\author{Keiichiro \textsc{Imura}$^{1,2}\thanks{E-mail: imura.keiichiro.j9@f.mail.nagoya-u.ac.jp}$, Yuki \textsc{Yoneyama}$^{2}$, Hideyuki \textsc{Ando}$^{2}$, Noriyuki \textsc{Kabeya}$^{3}$, Hitoshi \textsc{Yamaoka}$^{4}$, Nozomu \textsc{Hiraoka}$^{5}$, Hirofumi \textsc{Ishii}$^{5}$, Tsutomu \textsc{Ishimasa}$^{6}$ and Noriaki K. \textsc{Sato}$^{2,7}$}

\inst{$^{1}$Institute of Liberal Arts and Sciences, Nagoya University, Nagoya 464-8601, Japan\\
$^{2}$Graduate School of Science, Nagoya University, Nagoya 464-8602, Japan\\
$^{3}$Graduate School of Science, Tohoku University, Sendai 980-8578, Japan\\
$^{4}$RIKEN SPring-8 Center, Sayo, Hyogo 679-5148, Japan\\
$^{5}$National Synchrotron Radiation Research Center, Hsinchu 30076, Taiwan\\
$^{6}$Graduate School of Engineering, Hokkaido University, Sapporo 060-8628, Japan\\
$^{7}$Center for General Education, Aichi Institute of Technology, Toyota 470-0392, Japan
}

\abst{
We have synthesized new Tsai-type Yb-based intermediate-valence approximant crystals (ACs) with different degree of approximation to quasicrystal, Zn--Au--Yb 1/1 and 2/1 AC, and studied the external pressure effect on their Yb mean-valence $\nu$.
Whereas 1/1 AC distinctly exhibits a first-order-like jump in $\nu$ at a transition pressure $P_{\rm v}$, 2/1 AC only shows an indistinct anomaly at $P_{\rm v}$.
We have also studied the pressure dependence of the $\nu$ of Au--Al--Yb 1/1 AC, which is a prototypal AC exhibiting pressure-induced quantum criticality.
It shows a continuous valence anomaly at a critical pressure $P_{\rm c}$ where the magnetic susceptibility diverges toward zero temperature, in contrast to the valence jump in the Zn--Au--Yb 1/1 AC.
These results are discussed based on a theoretical model of quantum critical valence fluctuation.
}

\begin{document}
\maketitle

Quasicrystals (QCs) are long-range ordered materials featuring a symmetry incompatible with translational invariance.~\cite{Shechtman_1984,Levine_1984}
They have received a revived interest since the observation of quantum critical behaviors of Yb-based QCs with the intermediate valence (IV): the uniform magnetic susceptibility of Au--Al--Yb QC exhibits a power-law divergence towards zero temperature, $\chi(T) \sim T^{-0.51}$, at both ambient and high pressures.~\cite{Ishimasa_2011, Watanuki_2012, Deguchi_2012, Sato_2022}
In contrast, the relevant approximant crystal (AC) only shows the quantum critical feature at a critical pressure $P_{\rm c}$ ($\simeq 2$ GPa):~\cite{Matsukawa_2016} at other pressures, it shows Fermi liquid feature like a heavy fermion crystal.
Here, ACs are solids that possess similar local structures (clusters) as QCs but form a periodic array of clusters.~\cite{Elser_1985}
This striking difference in the $\chi(T)$ between QC and AC can be ascribed to the difference in the array of the clusters.

An intriguing structure--property relationship between the Yb mean-valence $\nu$ and the lattice constant was revealed in the Au--Al--Yb system and its substitution system Au$_{1-x}$Cu$_x$--Al$_{1-y}$Ga$_y$--Yb.
Note that whereas AC is characterized by three dimensional (3D) lattice constant $a_{\rm 3D}$ as usual, QC has periodicity in 6D space and is characterized by the 6D lattice constant $a_{\rm 6D}$.~\cite{Sato_2022}
When the $a_{\rm 3D}$ or $a_{\rm 6D}$ is changed by controlling
the composition $x$ and $y$, valence and magnetic singularities concomitantly occur at a critical lattice constant $a_{\rm c}$.~\cite{Imura_2020}
For the Au--Al--Yb 1/1 AC, Watanabe \& Miyake calculated $\nu$ as a function of $a_{\rm 3D}$ based on the quantum critical valence fluctuation model and succeeded in reproducing the anomaly at the $a_{\rm c}$.~\cite{Watanabe_2010, Watanabe_2018}
Although such a calculation has never been performed for QC owing to the absence of periodicity in QC structure, it is considered from the agreement between theory and experiment that quantum critical valence fluctuation is responsible for the quantum criticality of the Au--Al--Yb AC and QC.

\begin{figure}[t]
\begin{center}
\includegraphics[width=6cm]{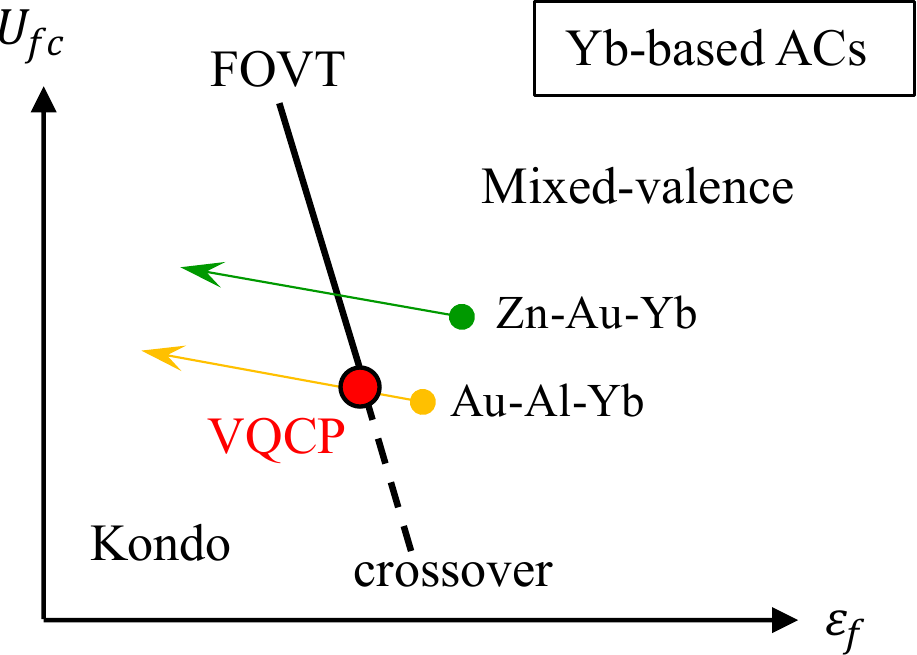}
\end{center}
\caption{(Color online)
Schematic $U_{fc}-\varepsilon_f$ phase diagram based on the quantum critical valence fluctuation model. $U_{fc}$ and $\varepsilon_f$ indicate Coulomb repulsion between $f$ and conduction electrons and $f$-electron energy, respectively. The closed circle, solid and dashed lines indicate valence quantum critical point (VQCP), first-order valence transition (FOVT) and crossover lines, respectively. The arrows indicate pressure evolution in Zn--Au--Yb and Au--Al--Yb ACs.
}
\end{figure}

According to the quantum critical valence fluctuation model, there is a first-order valence transition (FOVT) line on the phase diagram at $T=0$, see Fig.~1.~\cite{Watanabe_2011}
This FOVT line separates mixed-valence and Kondo regimes and terminates at the valence quantum critical point (VQCP).
The Au--Al--Yb 1/1 AC is in the mixed-valence regime at ambient pressure and is expected to move along the trajectory with increasing pressure as illustrated in the figure: when the external pressure is tuned to $P_{\rm c}$, the AC will be just at the VQCP.
Experimentally, the pressure-induced quantum criticality in the Au--Al--Yb 1/1 AC was confirmed by the observation of diverging feature in the $\chi(T)$ as mentioned above: however, it is not known if the Yb mean-valence $\nu$ shows a singular feature at $P_{\rm c}$.
It is also unknown if there is a system that shows the FOVT.
Furthermore, it remains an open question if the phase diagram characterized by FOVT and VQCP depends on the degree of approximation to QC: in the limit of infinite degree of approximation, i.e., QC, the VQCP would be transformed into a quantum critical phase.~\cite{Sato_2022}
To answer these questions, we have studied external pressure effect on the prototypal Au--Al--Yb 1/1 AC as well as a new IV system of Zn--Au--Yb ACs.
Note that the Zn--Au--Yb system has not only 1/1 phase but also 2/1 phase.~\cite{Ishimasa_2019}
Here we report the following findings: First, the $\nu$ of the Au--Al--Yb 1/1 AC shows anomaly at $P_{\rm c}$ where quantum critical behavior was observed in the susceptibility.
Second, the Zn--Au--Yb 1/1 AC shows a FOVT at a valence transition pressure $P_{\rm v} \sim 4.5$ GPa.
Finally, the Zn--Au--Yb 2/1 AC also shows a valence anomaly at $P_{\rm v} \sim 5.5$ GPa, however the anomaly is very obscured in contrast to the clear jump at $P_{\rm v}$ in the 1/1 AC.

We have synthesized Zn$_{100-x-y}$Au$_x$Yb$_y$ polycrystalline samples from high-purity starting materials of 4N(99.99\%)-Zn, 4N-Au, and 3N- or 4N-Yb.
Nominal compositions of these phases are $(x, y) = (4.5, 14.5)$ for 1/1 AC and $(x, y) = (9.0, 15.0)$ for 2/1 AC, respectively.
These materials were put in an aluminum crucible, and sealed in a quartz tube under high-vacuum of approximately $1.3\times 10^{-4}$ Pa.
The detailed melting process is described elsewhere~\cite{Ishimasa_2019}.
For 1/1 AC, after the melting process, heat treatment was made at 465$^\circ$C for 50 hs.
To confirm monophasic synthesis, we carried out powder x-ray diffraction (XRD) measurements using a laboratory XRD system (Rigaku, RINT2200) using the Cu $K_\alpha$ radiation.
The software package RIETAN-2000 was used for Rietveld refinement.~\cite{Izumi_2000}
Yb mean-valence measurements at room temperature were carried out under high pressure using high-energy-resolution fluorescence-detected x-ray absorption spectroscopy (HERFD-XAS) technique at the BL12XU inelastic x-ray scattering end station at SPring-8~\cite{Hamalainen_1991, Rueff_2010}.
Samples were pressurized using a He-gas controlled membrane-type diamond-anvil-cell with a culet diameter of 0.3 mm.
Silicone-oil was used as the pressure transmitting medium and the applied pressure was estimated from the shift of the ruby $R_1$ fluorescence line~\cite{Mao_1978, Yamaoka_2012}.
The detailed setup and analytical procedures such as spectrum fitting are described elsewhere~\cite{Imura_2020}.
The uniform magnetic susceptibility was measured using a commercial magnetometer (Quantum Design, MPMS) in the temperature range of 1.8 to 300 K.

\begin{figure}[tb]
\begin{center}
\includegraphics[width=8.5cm]{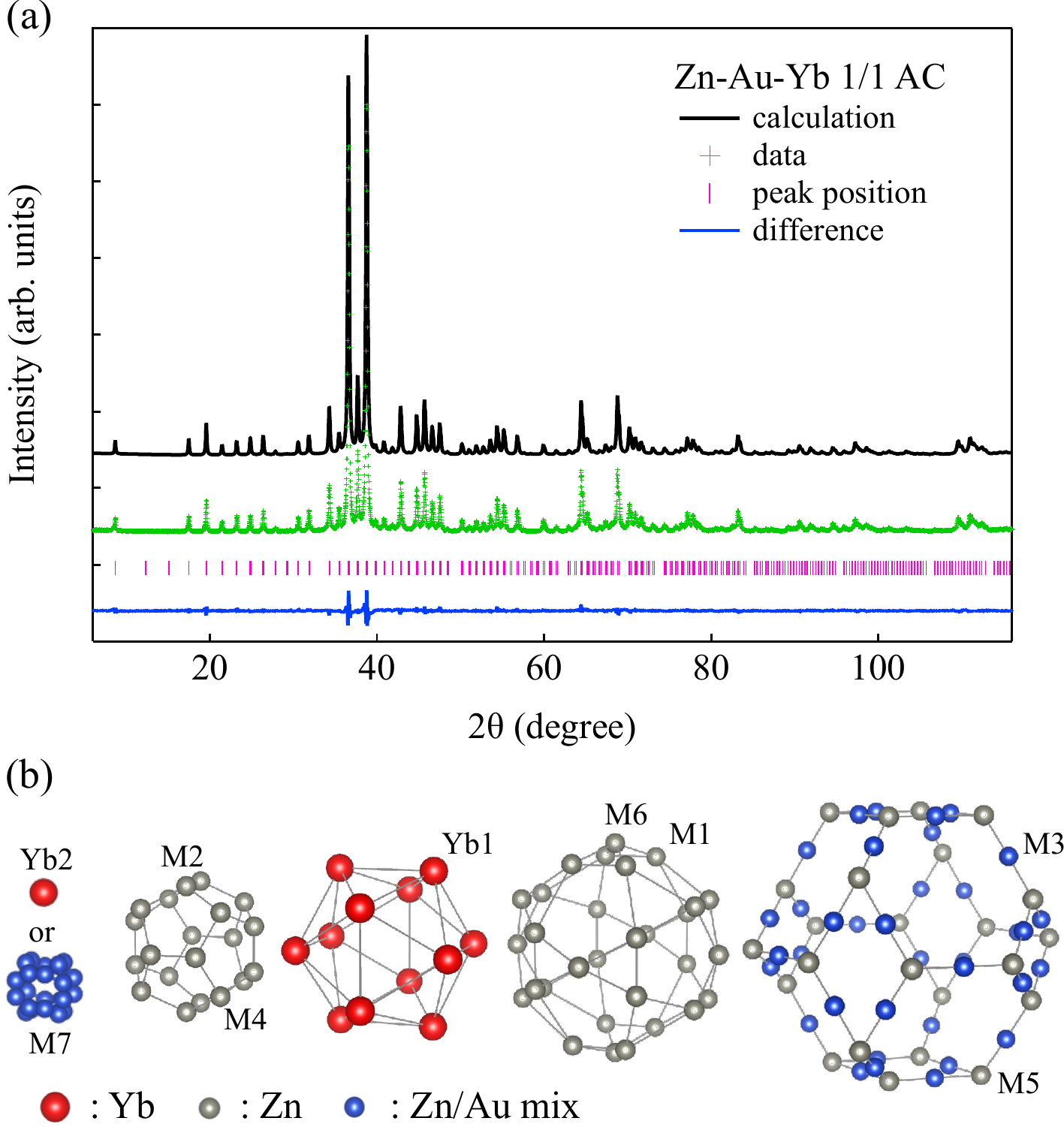}
\end{center}
\caption{(Color online)
(a) Powder x-ray diffraction pattern of Zn$_{81}$Au$_{4.5}$Yb$_{14.5}$ 1/1 AC. Calculated intensity, measured intensity, peak positions, and deviations are shown. (b) Structure model of cubic 1/1 AC of Zn--Au--Yb derived using Rietveld refinement. The cluster center contains a Yb ion (referred to as Yb2) or a tetrahedron composed of Zn/Au.
}
\end{figure}

\begin{table}[b]
\caption{Parameter set in the structure model of Zn--Au--Yb 1/1 AC.}
\label{table:data_type}
\centering
\begin{tabular}{ccccccc}
\hline
Site & Atom            & Set   & $x$       & $y$       & $z$       & $B$($\AA^2$) \\
\hline \hline
M1   & 0.98Zn+0.02Au   & 48$h$ & 0.3450(2) & 0.1944(2) & 0.1051(2) & 1.1(1)       \\
M2   & 0.92Zn+0.08Au   & 24$g$ & 0         & 0.2421(3) & 0.0880(3) & 1.6(1)       \\
M3   & 0.77Zn+0.23Au   & 24$g$ & 0         & 0.5952(2) & 0.6506(2) & 0.18(7)      \\
M4   & 0.48Zn          & 16$f$ & 0.158(2)  & 0.158(2)  & 0.158(2)  & 1.5(5)       \\
M4'  & 0.52Zn          & 16$f$ & 0.139(2)  & 0.139(2)  & 0.139(2)  & 0.6(5)       \\
M5   & Zn              & 12$e$ & 0.1940(3) & 0         & 0.5       & 0.56(8)      \\
M6   & Zn              & 12$e$ & 0.4108(5) & 0         & 0         & 1.8(1)       \\
M7   & 0.064Zn+0.019Au & 48$h$ & 0.032(2)  & 0.067(2)  & 0.097(2)  & 3(1)         \\
Yb1  & Yb              & 24$g$ & 0         & 0.1898(1) & 0.3005(1) & 0.46(3)      \\
Yb2  & 0.51Yb          & 2$a$  & 0         & 0         & 0         & 0.2(4)       \\
\hline
\end{tabular}
\end{table}

Figure 2(a) shows the powder XRD pattern of Zn--Au--Yb 1/1 AC.
Rietveld refinement is successfully converged by assuming the space group of $Im\overline{3}$.
The parameter set is summarized in Table I.
The unit cell is a body-centered cubic (BCC) with a lattice parameter of $a_{1/1} =$ 14.300(2) $\AA$.
The obtained structure model of Zn--Au--Yb 1/1 AC is drawn in Fig. 2(b), where a Tsai-type cluster with three polyhedral shells is placed in the BCC framework.
In contrast to Au--Al--Yb 1/1 AC~\cite{Ishimasa_2011}, the cluster centers contain a single Yb (Yb2) atom or tetrahedron consisting of Zn/Au mixed site (M7) with a probability of 51:49.
The reliable factor, $R_{\rm wp}$, is 6.9\% in this model.
The fact that structure models without Yb2 site always showed $R_{\rm wp}$ greater than 8\% demonstrates the validity of the current model with Yb atom at Yb2 site.

The icosahedron of the second shell consisting of Yb (red ball) atoms is sandwiched between the dodecahedron of the first shell (consisting of M2 and M4 sites) and the icosidodecahedron of the third shell (consisting of M1 and M6 sites).
The rhombic triacontahedron composed of M3 and M5 sites forms a network by wrapping around these shells.
The 2/1 AC is a cubic structure with lattice parameter $a =$ 23.271(2) $\AA$ and space group $Pa\bar{3}$.
Similar to 1/1 AC, this structure also consists basically of Tsai-type clusters, but due to complexity of the unit cell containing about 700 atoms, the details of the structure have not been clarified.~\cite{Ishimasa_2019}   

\begin{figure}[b]
	\begin{center}
	\includegraphics[width=8.5cm]{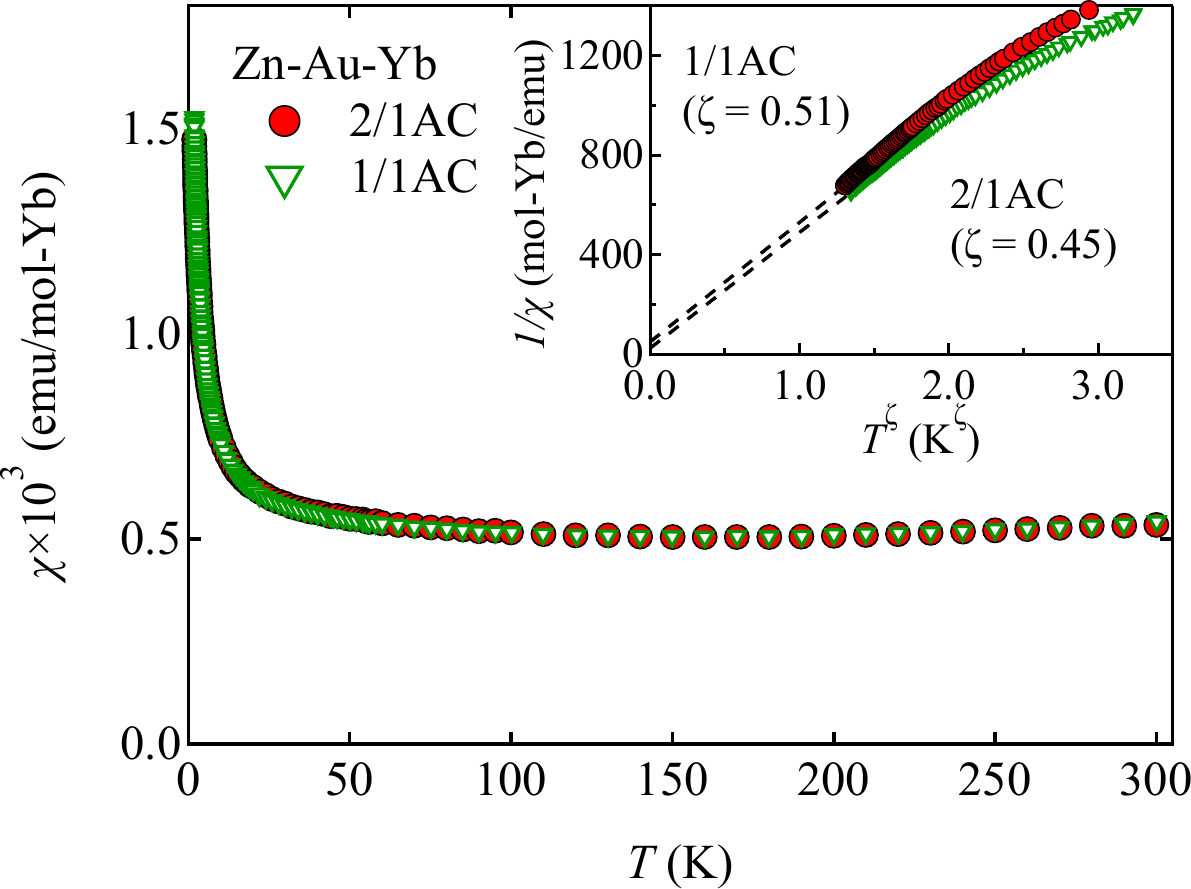}
	\end{center}
	\caption{(Color online)
	(a) Temperature dependence of uniform magnetic susceptibility $\chi(T)$ of Zn--Au--Yb 1/1 AC and 2/1 AC. (Inset) Inverse magnetic susceptibility plotted against $T^{\zeta}$ with $\zeta=0.51$ and 0.45 for Zn--Au--Yb 1/1 AC and 2/1 AC, respectively.
	}
	\end{figure}

Figure 3 shows the temperature dependence of the magnetic susceptibility $\chi(T)$ of Zn--Au--Yb 1/1 and 2/1 AC measured at an external field of $H =$ 1000 Oe.
Note that although an anomaly originating from an antiferromagnetic order of Yb$_2$O$_3$ is often observed at around 2 K,~\cite{Yoneyama_MC} there is no trace of impurity phase in the figure.
This indicates the high quality of the samples studied here.
We note that the $\chi$ of the Zn--Au--Yb ACs is more than one order of magnitude smaller than that of Au--Al--Yb 1/1 AC, consistent with the fact that the Yb mean-valence of the Zn--Au--Yb ACs is smaller than that of the Au--Al--Yb 1/1 AC (see below).
In the inset, the inverse of the susceptibility is plotted against $T^\zeta$, where the exponent of the 1/1 and 2/1 AC is $\zeta=0.51$ and 0.45, respectively.
Although we need lower temperature experiment to accurately determine the critical behavior toward zero temperature, the straight-line extrapolation with the above exponent seems not to go through the origin of the figure, similar to the ambient-pressure Au--Al--Yb 1/1 AC, indicating no divergence of the susceptibility.

\begin{figure*}[t]
	\begin{center}
	\includegraphics[width=15cm]{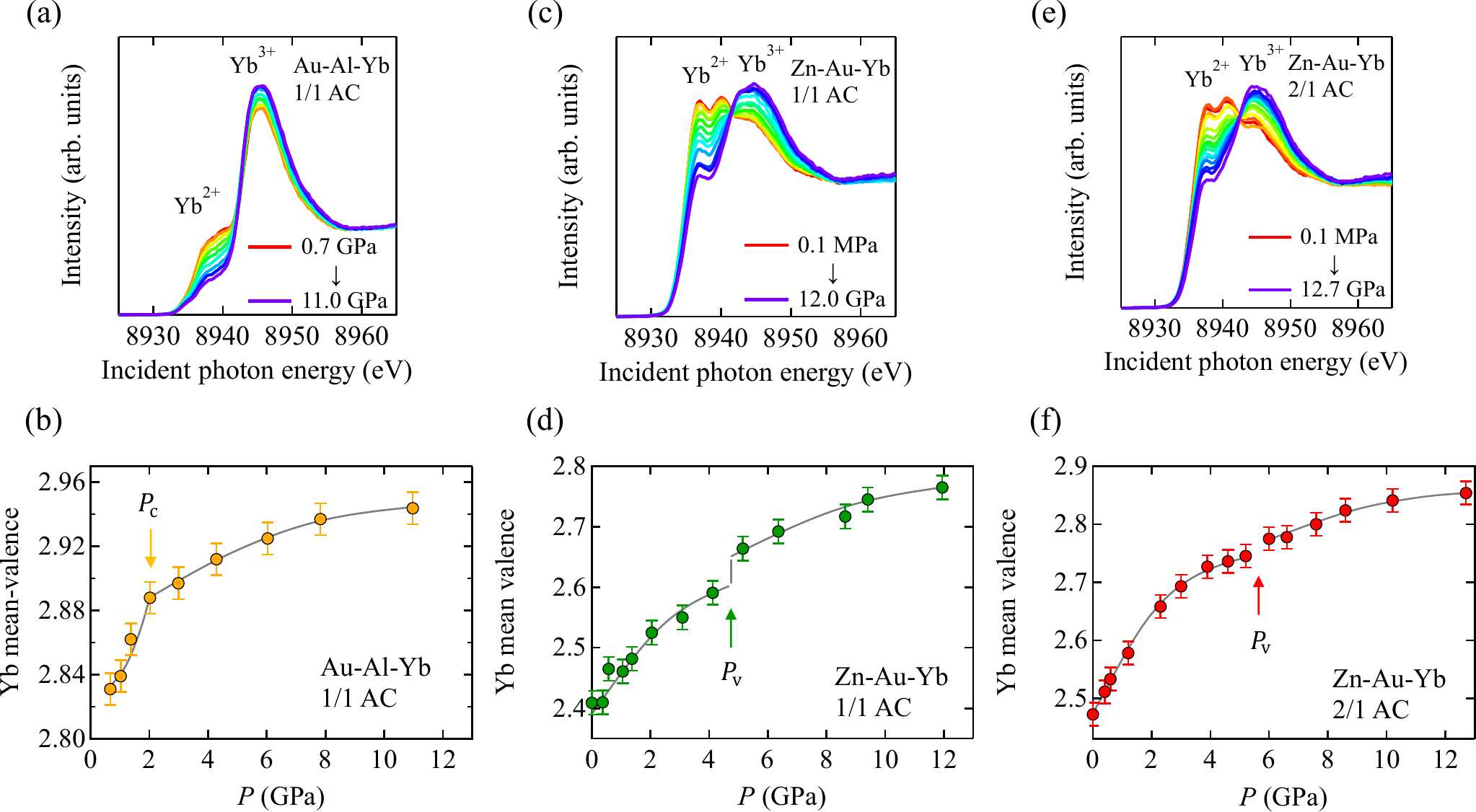}
	\end{center}
	\caption{(Color online)
	Pressure evolution of incident photon energy dependence of HERFD-XAS of (a) Au--Al--Yb 1/1 AC up to 11.0 GPa, (c) Zn--Au--Yb 1/1 AC up to 12.0 GPa, and (e) Zn--Au--Yb 2/1 AC up to 12.7 GPa. All measurements were performed at 300 K. We find a characteristic common to all spectra: the lower-energy structure at around 8939 eV decreases with increasing pressure, whereas the higher-energy structure at around 8945 eV increases with pressure. Pressure dependence of Yb mean-valence of (b) Au--Al--Yb 1/1 AC, (d) Zn--Au--Yb 1/1 AC, and (f) Zn--Au--Yb 2/1 AC. $P_{\rm c}$ in Au--Al--Yb 1/1 AC is the critical pressure at which the magnetic susceptibility showed quantum critical behavior, and $P_{\rm v}$ in Zn--Au--Yb ACs is a valence transition pressure. The solid lines are guide to the eye.
	}
	\end{figure*}

Pressure evolution of the Yb-$L_{\rm III}$ HERFD-XAS spectra of the IV ACs is shown in Fig.~4.
First we study Au--Al--Yb 1/1 AC [see Fig.~4(a)].
At 0.7 GPa (the lowest pressure in this experiment), we observe characteristic two-peak structure of the IV system: the peak at 8939 and 8945 eV originates from Yb$^{2+}$ and Yb$^{3+}$ state, respectively.
Using the definition of Yb mean-valence $\nu = I_3/(I_2+I_3) +2$ (where $I_2$ and $I_3$ are integrated intensities of Yb$^{2+}$ and Yb$^{3+}$ configuration, respectively), we obtain $\nu =$ 2.83, which is slightly larger than the ambient-pressure value of $\nu =$ 2.78.~\cite{Imura_2020}
When the external pressure is increased, the lower-energy structure decreases in intensity, whereas the higher-energy structure increases: as a result, the $\nu$ increases with pressure, consistent with the fact that Yb$^{3+}$ is smaller in size than Yb$^{2+}$ and hence preferred at high pressures.

Pressure dependence of Yb mean-valence $\nu(P)$ in Au--Al--Yb 1/1 AC is displayed in Fig.~4(b).
The arrow indicates the critical pressure $P_{\rm c}$ ($\simeq$ 2 GPa) at which the magnetic susceptibility diverges toward zero temperature.~\cite{Matsukawa_2016}
We find that the $\nu$ shows anomaly at around $P_{\rm c}$. This answers the first question mentioned above: the Yb mean-valence and the susceptibility exhibit the anomaly at the same critical pressure $P_{\rm c}$.

Figure 4(c) and (e) show pressure evolution of the Yb-$L_{\rm III}$ HERFD-XAS spectra of Zn--Au--Yb 1/1 and 2/1 AC, respectively.
The overall feature of the spectra is similar to that of the Au--Al--Yb AC, indicating that the Zn--Au--Yb 1/1 and 2/1 ACs are both a new family of IV-ACs.
It seems that the lower-energy structure at around 8939 eV is split into two peaks at 8938 and 8940.5 eV.
We note that both peaks decrease in intensity with increasing pressure, which allows us to assign them as the Yb$^{2+}$ component.
Assuming this assignment, we evaluate the ambient-pressure Yb mean-valence of the Zn--Au--Yb system as $\nu=$ 2.41 and 2.47 for 1/1 and 2/1 AC, respectively.
These are smaller than that of Au--Al--Yb AC, consistent with the above-mentioned result of magnetic susceptibility.
When the external pressure is increased, the $\nu$ increases as expected.

Pressure dependence of Yb mean-valence $\nu(P)$ in Zn--Au--Yb ACs are summarized in Figs.~4(d) and (e). 
In 1/1 AC [see Fig.~4(d)], $\nu$ increases monotonically with pressure up to $P_{\rm v} \sim$ 4.5 GPa.
We stress that the $\nu$ changes discontinuously from 2.60 to 2.65 at the pressure, suggesting the emergence of the FOVT.
When the pressure is further increased, the $\nu$ increases monotonically again, finally reaching approximately 2.78 at $P =$ 12.0 GPa.
This observation answers the second question mentioned above: there exists IV AC that shows the FOVT.
In contrast, the 2/1 AC only shows an obscured anomaly (a small jump of $\nu =$ 2.75 to 2.78 or just a crossover) at $P_{\rm v} \sim$ 5.5 GPa [see Fig.~4(e)].
When the pressure is further increased, the $\nu$ increases monotonically similar to the 1/1 AC, reaching approximately 2.85 at 12.0 GPa.
From this finding, we suggest that the valence transition at $P_{\rm v}$ becomes obscured when the approximation degree increases.
This answers the final question mentioned above: the valence anomaly diminishes (or broadens) with increasing the degree of approximation to QC.

We note the difference in spectral shape between the Au--Al--Yb and Zn--Au--Yb 1/1 ACs: it may be related to the difference in the crystal structure on the analogy of the fact that in Au--Ge--Yb 1/1 AC, which possesses two Yb-atom sites as seen in Fig. 2(b), two Yb ions have different valence states from each other.~\cite{Matsunami_2017}
It would be interesting to confirm that the presence of the crystallographically different sites for the Yb atoms is responsible for the splitting of the lower-energy peak.

Finally, we attempt to interpret the present experimental findings with the help of the theoretical phase diagram schematically illustrated in Fig.~1.~\cite{Watanabe_2011}
From the agreement between experiment and theory (see the above introductory part), it is inferred that the Au--Al--Yb 1/1 AC moves along the (online yellow) line via the VQCP with increasing pressure.
In contrast, the Zn--Au--Yb ACs cross the FOVT line when they move along the (online green) trajectory. 
Here, we assumed that since there is no visible difference in the ambient-pressure magnetic susceptibility between the Zn--Au--Yb 1/1 and 2/1 ACs, they start from the same ''initial point'' (on the phase diagram) corresponding to $P=0$.
For the difference in the valence anomaly at the $P_{\rm v}$ between the Zn--Au--Yb 1/1 and 2/1 ACs, we consider that the phase diagram characterized by the FOVT and VQCP will vary with the approximation degree.

In summary, we synthesized Zn--Au--Yb 1/1 and 2/1 ACs.
The Rietveld analysis for the 1/1 AC showed that Yb ions are at not only the vertices of icosahedron but also the center of Tsai-type cluster.
HERFD-XAS experiments showed that both Zn--Au--Yb 1/1 and 2/1 ACs are in IV states and exhibit an anomaly in the $\nu(P)$ curve at the valence transition pressure $P_{\rm v}$. It should be stressed that the valence anomaly of the 1/1 AC is clearly accompanied by a first-order-transition-like jump, whereas that of the 2/1 AC is obscured.
This suggests that the phase diagram characterized by the FOVT and VQCP will vary with the approximation degree to QC.
We also studied the pressure effect on the Yb mean-valence of Au--Al--Yb 1/1 AC and found that an anomaly emerges at the critical pressure $P_{\rm c} \simeq$ 2 GPa where the magnetic susceptibility shows quantum critical feature. 
It is stressed that the feature of anomaly in this prototypal IV AC differs from that in the newly found Zn--Au--Yb ACs.
Comparing these two systems, we discussed that when the external pressure is increased, the Au--Al--Yb and Zn--Au--Yb ACs pass through different trajectories on the theoretical phase diagram.

\begin{acknowledgment}
This work was supported by JSPS KAKENHI (Grant Numbers 21H01028 and 22H04591).The synchrotron radiation experiments were performed at the NSRRC Taiwan beamline BL12XU, SPring-8 (under SPring-8 Proposal Nos. 2021B4252, 2022A4257, and 2022B4265, corresponding NSRRC Proposal No. 2021-2-314).
\end{acknowledgment}

\end{document}